\documentclass[twocolumn,showpacs,superscriptaddress,preprintnumbers,amsmath,amssymb]{revtex4-1}
\usepackage{graphicx}
\usepackage{dcolumn}
\usepackage{appendix}
\usepackage{bm}
\usepackage{enumitem}
\usepackage{booktabs}

\usepackage[usenames,dvipsnames]{xcolor}
\usepackage{subfigure}
\usepackage{bbm}
\usepackage{enumerate}
\usepackage[
  bookmarks=true,
  colorlinks,
  linkcolor=black,
  urlcolor=black,
  citecolor=black,
  plainpages=false,
  pdfpagelabels,
  final,
  breaklinks=true
]{hyperref}

\usepackage{titlesec}
\usepackage{array}
\usepackage{dsfont}
\usepackage{physics}
\usepackage{multirow}
\usepackage{pifont}
\usepackage{soul}

\begin{document}

\title{Quantum phenomena in attosecond science}

\author{Lidice Cruz-Rodriguez}
\email{lidice.rodriguez@ucl.ac.uk}
\affiliation{Department of Physics and Astronomy, University College London, London WC1E 6BT, UK}

\author{Diptesh Dey}
\email{diptesh.dey@northwestern.edu}
\affiliation{Department of Chemistry, Northwestern University, Evanston, Illinois 60208, USA}

\author{Antonia Freibert}
\email{afreiber@physnet.uni-hamburg.de}
\affiliation{Institute for Nanostructure and Solid-State Physics, Department of Physics and Center for Free-Electron Laser Science, University of Hamburg, 22761 Hamburg, Germany}

\author{Philipp Stammer}
\email{philipp.stammer@icfo.eu}
\affiliation{ICFO -- Institut de Ciencies Fotoniques, The Barcelona Institute of Science and Technology, 08860 Castelldefels (Barcelona), Spain}
\affiliation{Atominstitut, Technische Universität Wien, 1020 Vienna, Austria}

\date{\today}

\begin{abstract}
Attosecond science has opened up new frontiers in our understanding of processes happening on the intrinsic timescale of electrons. The ability to manipulate and observe phenomena at the attosecond level has yielded groundbreaking insights into processes such as electron dynamics and the behavior of matter under extreme conditions. This interdisciplinary field bridges various research areas such as quantum optics, quantum chemistry and quantum information science facilitating a cohesive understanding. However, despite many emerging successful applications, the discussion about intrinsic quantum effects has mainly been ignored.
In this Perspective, we explore the latest advancements in quantum phenomena within attosecond science, encompassing both experimental and theoretical progress. Specifically, in the context of high-harmonic generation and above-threshold ionization, we focus on discerning genuinely quantum observations and distinguishing them from classical phenomena. Additionally, we illuminate the often overlooked yet significant role of entanglement in attosecond processes, elucidating its influence on experimental outcomes.

\end{abstract}

\maketitle

\section{\label{sec:intro}Introduction}

\begin{figure*}
    \centering
     \includegraphics[scale=0.54]{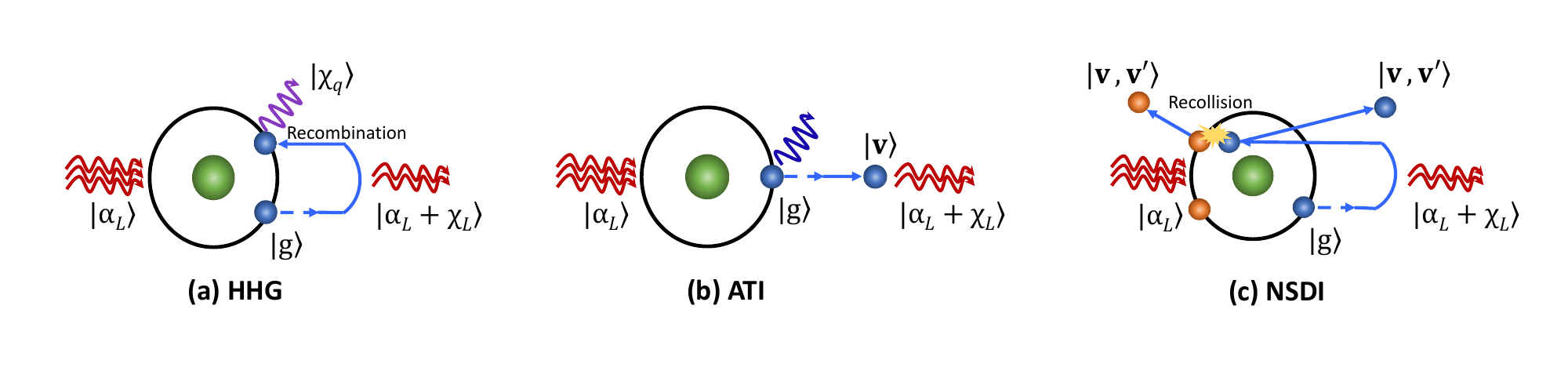}
      \caption{An illustration of the key strong-field processes: (a) high-harmonic generation (HHG), (b) above-threshold ionization (ATI) and (c) nonsequential double ionization (NSDI) along with the corresponding quantum states of light and matter. The states of the driving field mode before and after interaction are represented by $\ket{\alpha_{L}}$ and $\ket{\alpha_{L} + \chi_{L}}$, respectively. The initial ground state and final continuum stateof the electron are given by $\ket{g}$ and $\ket{\vb{v}}$. High-order harmonic modes, due to HHG, are denoted by $\ket{\chi_{q}}$. }
  \label{fig:illustration}
\end{figure*}

For over a century, investigating matter-field interactions has been a fundamental aspect of physics and chemistry research. Matter exposed to high-intensity laser sources, where the field strength is comparable to the binding energy of the atom or molecule, undergo fascinating non-linear phenomena only accessible in the strong-field regime. The interaction with strong-laser fields leads to photoionization, producing photoelectrons via the above-threshold ionization (ATI) or high-order ATI (HATI) mechanisms~\cite{agostini1979free,becker2002above}. Furthermore, the laser-driven electron can return to the parent ion, emitting high-order harmonics (HHG) \cite{ferray1988multiple} or producing a second photoelectron via the nonsequential double ionization process (NSDI) \cite{l1983multiply}. 
An overview of these fundamental strong-field processes is illustrated in Fig. \ref{fig:illustration}. 

The observation of the HHG process led to the subsequent development of extremely short-laser pulses of only a few hundred attoseconds ($1 \, \rm attosecond =10^{-18} \rm s$) \cite{PaulScience01,hentschel2001attosecond}, creating an entirely new field of research known as attosecond science \cite{CorkumNaturePhys07}. Within this field, novel experimental techniques have allowed us to look at the electron dynamics in their natural timescale with impact in atomic, molecular, and solid-state physics \cite{calegari2016advances}. 

The development of novel theoretical approaches has accompanied the advancement of experimental observations.
A general challenge faced by attosecond investigations is the presence of many degrees of freedom with complex and, in general, unknown interactions. 
It is generally accepted that the field-matter interaction is described by the time-dependent Schrödinger equation (TDSE). However, in typical strong-field scenarios, the exact solution of the TDSE is challenging and becomes unfeasible as the number of degrees of freedom increases. Additionally, the physical interpretation of the results from ab initio methods is non-trivial~\cite{armstrong2021dialogue}. 
Henceforth, approximate methods have been constantly developed to describe strong-field matter interaction, usually by treating the light field classically \cite{stammer2023limitations}. The most widespread method is the strong-field approximation (SFA) \cite{amini2019symphony}, which has been very successful in modeling strong-field induced phenomena. However, as attosecond science extends to more complex systems, the general quest for novel theoretical approaches to account for electron-electron correlations or the coupling between the electron and nuclear degrees of freedom is crucial.
Also closely related to the theoretical modeling of strong-field phenomena has been the question of the quantum or classical origin of the experimental observations. For example, in the process of NSDI, the role of electron-electron correlation has been long known \cite{becker2012theories}, but whether we are in the presence of classical \cite{HoPRL05} or quantum \cite{hao2014quantum,maxwell2016controlling,quan2017quantum} correlations has remained an open question. 

Entanglement represents one of the most iconic departures from classical physics. Most of the laser-induced phenomena involve the breaking of an originally bound subsystem, leading to the creation of two or more subsystems that can in general be entangled. However, over the years, the role of entanglement in strong-field phenomena has mainly remained unexplored. Only recently, new studies have focused on understanding the role of entanglement in the photoionization process; both looking at the entanglement between the photoelectron and the parent ion \cite{Vrakking2022,koll2022experimental,Nishi2019,Busto2022,ruberti2023bell,Nabekawa2023}, the electron-electron entanglement in the NSDI process \cite{christov2019phase,maxwell2022entanglement} and the entanglement between electronic and nuclear degrees of freedom\cite{McKemmish2011,Vatasescu2013,Vatasescu2015,Izmaylov2017,SanzVicario2017} and its impact on ultrafast photochemistry \cite{Blavier2022}.

Additionally, new approaches have been developed to achieve a full quantum optical description of attosecond processes, including the matter and the light field. Over the years, the theoretical models employed to describe strong-laser field phenomena have relied on a quantum description of matter while considering a classical laser field. Only recently, the bridge between strong-field physics and quantum optics has been built by developing a fully quantum description of the laser-matter interaction \cite{gorlach2020quantum, lewenstein2021generation, stammer2023quantum}. The HHG and ATI processes have been studied within a complete quantum electrodynamics framework, showing the appearance of non-classical and entangled states of light. 
Furthermore, quantum optical approaches allow to consider non-classical states of light driving the processes~\cite{gorlach2023high, stammer2023role, even2023photon} and challenge the widespread semi-classical models~\cite{stammer2023limitations}.
These recent developments have opened the possibility of connecting attosecond physics with quantum information science and modern quantum technologies \cite{bhattacharya2023strong, lewenstein2022attosecond}.

In this Perspective, we address how the notions of entanglement and non-classical states of light emerge in processes on the attosecond time scale. The paper will be structured as follows. In section \ref{sec:1}, we will address the quantum phenomena of radiation in attoscience, focusing on the recently developed full quantum optical descriptions, and consider the HHG and ATI process under this new formulation. We will discuss the observables to witness non-classicality, as well as the appearance of entanglement of the field modes. In section \ref{sec:2}, we explore the recent analysis of the role of entanglement in attosecond processes, with emphasis on ion-photoelectron entanglement, nuclear-electronic entanglement, and electron-electron entanglement.
We conclude in section \ref{sec:conclusion} by providing an outlook where novel quantum phenomena in attosecond science could emerge.

\section{\label{sec:1}Quantum phenomena of radiation in attoscience}

In this section, we will focus on the two most prominent phenomena in attosecond science, namely light scattering and photoionization by means of the process of high-harmonic generation (HHG) and above-threshold ionization (ATI), respectively.
We particularly focus on the question which observations are genuinely quantum and how to distinguish non-classical processes from its classical counterparts. 
This is of particular current interest since the theoretical approaches for the aforementioned processes have recently developed from the semi-classical to the full quantum optical description \cite{gorlach2020quantum, lewenstein2021generation, stammer2023quantum}. 
The additional degrees of freedom from the quantization of the electromagnetic field allow to consider field observables not conceivable before, and using Hilbert space constructs for the driving field as well as for the generated harmonic radiation offer new possibilities to envision experiments for optical quantum technologies \cite{o2009photonic, gilchrist2004schrodinger}.

\subsection{High harmonic generation}

It is known that classical radiation is generated from oscillating charge currents \cite{scully1999quantum}. 
In the semi-classical description of the HHG process \cite{lewenstein1994theory} the electron is driven by an intense classical electromagnetic field $E_{cl}(t)$, which is coupled to the electrons dipole moment. 
The induced time-dependent dipole moment, is then analyzed to obtain the spectrum of the scattered light in the HHG process.
Here, the scattered light is again obtained from a classical oscillating charge current, emphasizing that the generated radiation is classical. 
However, within the semi-classical description of attosecond processes the oberservables of the light field is limited. In particular, field properties obtained from a quantum optical perspective remain elusive and genuine non-classical features of the field are hidden. 

We will first briefly review the existing work on the quantum optical description of radiation in attosecond processes, some of them going beyond classical radiation properties, and then elaborate possible new avenues to investigate the non-classical radiation properties in strongly driven systems. 

\subsubsection{Quantum optical formulation of HHG}

To construct a full quantum optical formulation of the process of HHG, it is inevitable to quantize the radiation field. A detailed quantum electrodynamical quantization procedure and derivation of the effective interaction Hamiltonian is given in \cite{stammer2023quantum}.
Constitutive in the full quantum description is the electric field operator $E_Q(t)$ coupled to the electron dipole moment, instead of the semi-classical coupling via the classical oscillating electric field $E_{cl}(t)$. 
Furthermore, quantization of the field requires to define the initial state boundary condition imposed by the experiment. Assuming the HHG process is driven by classical coherent laser light, described by a coherent state $\ket{\alpha_L}$, the classical interaction can be recovered \cite{lewenstein2021generation, stammer2023limitations, stammer2023quantum2}. 
This classical interaction leads to product coherent states in the harmonic field modes after the interaction 
\begin{align}
\label{eq:product_state}
    \ket{\{ 0_q \} } \xrightarrow[\text{HHG}]{} \ket{\{ \chi_q \} } = \bigotimes_q \ket{\chi_q} ,
\end{align}
where $\ket{\{0_q\}}$ is the initial vacuum state of all harmonics.
This finding is in agreement with the semi-classical picture, and induces a shift $\chi_L$ in the coherent state driving laser amplitude \cite{lewenstein2021generation}. The state after the interaction thus reads $\ket{\alpha_L + \chi_L}$ and accounts for the depletion of the pump field. 
In addition, the coherent state amplitudes of the harmonic field modes are given by the Fourier transform of the dipole moment expectation value, i.e. representing a classical oscillating charge current 
\begin{align}
\label{eq:displacement}
    \chi_q \propto \int_0^\infty dt \bra{g}d(t)\ket{g} e^{i \omega_q t},
\end{align}
which further manifests the validity of the semi-classical approach. 

However, when solving the interaction of the radiation field with the dipole moment of the electron, it is crucial to neglect the dipole moment correlations \cite{sundaram1990high, stammer2022theory} in order to obtain the product coherent states in \eqref{eq:product_state}. 
This is because the exact operation for the parametric process of HHG acting on the field state is given by 
\begin{align}
\label{eq:Kraus_exact}
    K_\mathrm{HHG} = \bra{g} \mathcal{T} \exp[- i \int_0^\infty dt \, d(t) E_Q(t)] \ket{g},
\end{align}
which is in general hard to solve since the commutator of the dipole moment at different times remains an operator in the electron Hilbert space $[d(t_1) , d(t_2) ] \in \mathcal{H}$, and therefore the different field modes in the electric field operator mix \cite{stammer2022high, stammer2022theory, stammer2023entanglement}.
Neglecting dipole moment correlations in \eqref{eq:Kraus_exact}, the operator can be solved and leads to a multimode displacement operation in phase-space 
\begin{align}
\label{eq:Kraus_approx}
    K_\mathrm{HHG} & \simeq \mathcal{T} \exp[- i \int_0^\infty dt \, \bra{g}d(t)\ket{g} E_Q(t)] \nonumber \\
    & = \prod_q D[\chi_q], 
\end{align}
where $D[\chi_q] = \exp[\chi_q b_q^\dagger - \chi_q^* b_q]$. Since the approximate operation in \eqref{eq:Kraus_approx} is linear in the field operators $b_q^{(\dagger)}$ the modes do not couple and the field remains in product coherent states as reported in \cite{lewenstein2021generation, rivera2022strong}.
Going beyond, and taking into account dipole moment correlations, can lead to interesting effects such as field correlations \cite{pizzi2023light}, mode squeezing \cite{gorlach2020quantum, stammer2023entanglement} or entanglement between the modes \cite{stammer2023entanglement, stammer2022theory}. Further investigation about the role of the aforementioned correlations could reveal novel radiation properties of the generated harmonics. 

\subsubsection{\label{sec:HHG_properties}Radiation properties: observables to witness non-classicality}

With the quantum optical formulation of the HHG process we can consider field observables beyond the HHG spectrum which is obtained from the oscillating classical charge current of the electron. 
Recent investigations have shown that there is a variety of interesting observations which further characterize the field properties. These allow to obtain insights about the associated quantum state of the field modes and to witness non-classical properties. We refer to Table \ref{table:HHG} for a collection of observables in HHG and a classification of their genuine quantum origin and possible classical counterparts. 

\begin{figure}
    \centering
	\includegraphics[width=0.49\columnwidth]{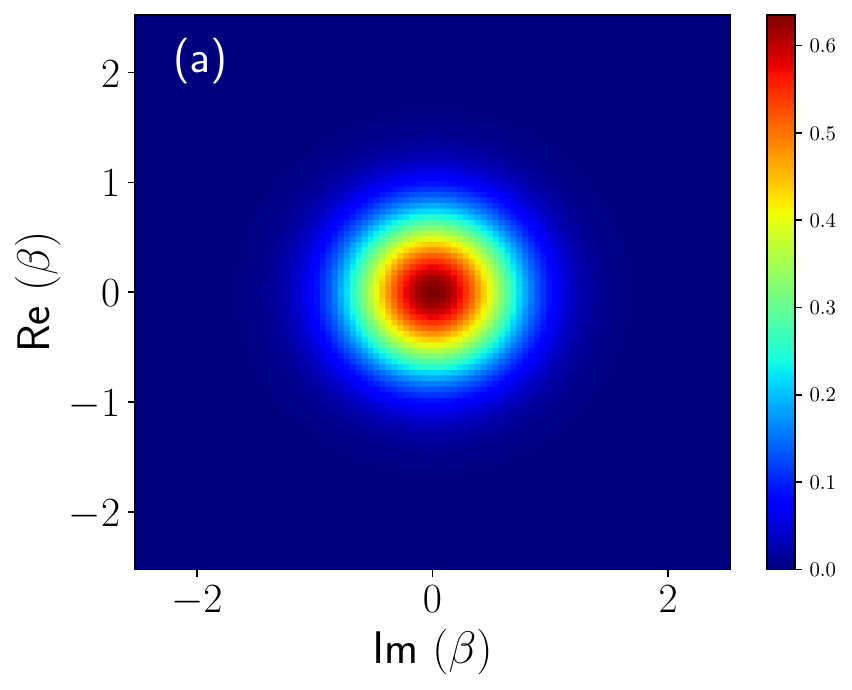}
	\includegraphics[width=0.49\columnwidth]{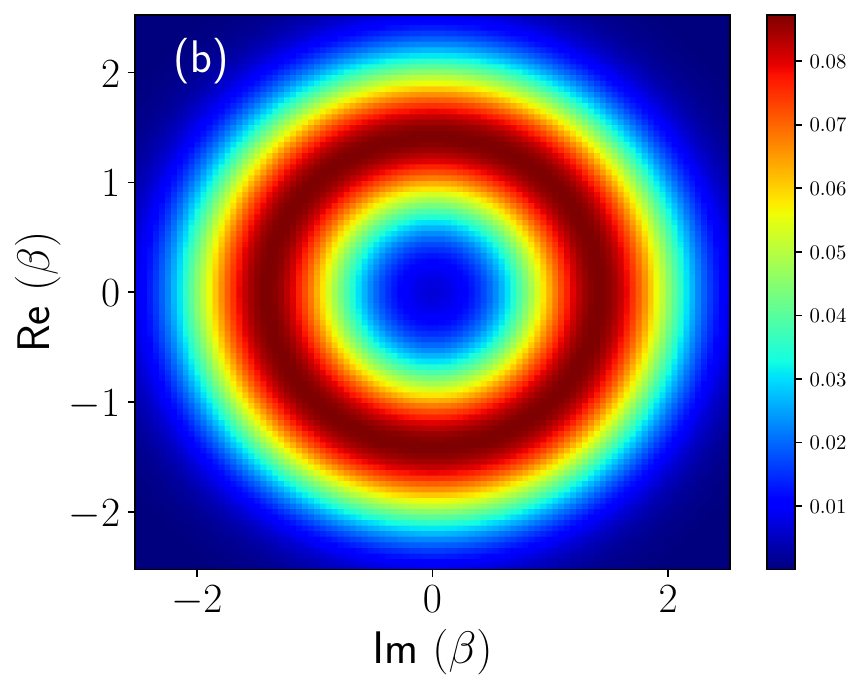}
	\includegraphics[width=0.49\columnwidth]{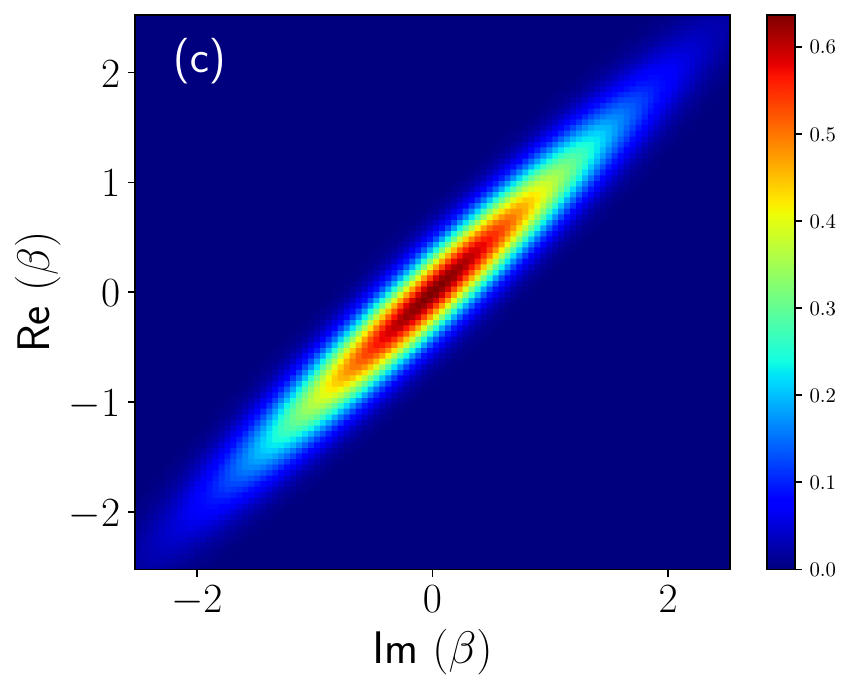}
 	\includegraphics[width=0.49\columnwidth]{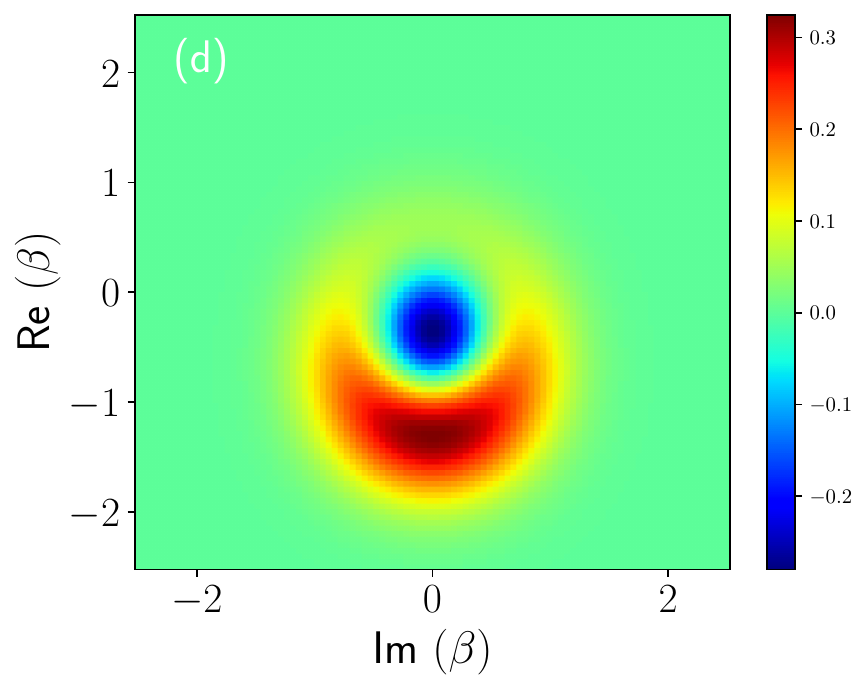}
	\caption{Examples of Wigner functions $W(\beta)$ for different states considered in the quantum optical formulation of HHG. The top row shows classical states: (a) coherent state $\ket{\alpha}$ and (b) phased mixed coherent state $\rho \propto \int_0^{2 \pi} d\phi \dyad{\abs{\alpha} e^{i \phi}}$. In the bottom row are non-classical states (c) squeezed vacuum and (d) optical cat state. }
  \label{fig:HHG_wigner}
\end{figure}

Beyond the spectrum, which has its classical counterpart in HHG, there are a variety of different observables accessible within the quantum optical description. 
One of the most prominent examples to witnesses non-classical field states is the Wigner function $W(\beta)$, obtained from homodyne measurements of the probe field together with a local oscillator \cite{gerry2023introductory, schleich2011quantum, rivera2021new}. 
Such homodyne measurements allow to measure the field quadrature of the probe field and to reconstruct it's Wigner function (for experimental details related to HHG see \cite{stammer2023quantum, bhattacharya2023strong}). 
In the quantum optical context the Wigner function is commonly used to classify the state into classical and non-classical fields. Since classical probability distributions are strictly positive, negativities in the Wigner function quasi-probability are witnesses of non-classical light \cite{schleich2011quantum, footnote1}. 
In the process of HHG, there has been recent progress in studying the radiation properties of the scattered light field using the Wigner function. 
For instance in \cite{pizzi2023light} it was shown that initial matter correlations are imprinted in the Wigner function, leading to non-Gaussian Wigner functions with $W(\beta) \ge 0$, and allows for the engineering of the quantum states of the harmonic field modes. 
Another direction to perform quantum state engineering of light using the process of HHG has been developed in \cite{lewenstein2021generation} and subsequent works \cite{rivera2022strong, stammer2022high, stammer2022theory, stammer2023quantum, lamprou2023nonlinear}. 
Instead of using correlated matter systems, the state engineering is approached with conditional measurements. Correlating the harmonic radiation with the driving field allows to generate optical cat and kitten states in the driving field mode \cite{lewenstein2021generation, rivera2022strong, stammer2022high} or the harmonic field modes \cite{stammer2022high, stammer2022theory}.
All those methods allow to generate high photon number non-classical field states ranging from the infrared to the extreme ultraviolet regime. 
The intensities of the non-classical optical cat states are even sufficient to drive non-linear processes \cite{lamprou2023nonlinear} and allow to perform quantum state engineering of the second harmonic field obtained when driven by a non-classical field. 
Another prominent example of non-classical field properties is the squeezing of a field quadrature. Squeezing allows to manipulate the field fluctuations along a specific quadrature, the most prominent being the amplitude or phase squeezing \cite{walls1983squeezed}. 
While the amplitude squeezing reduces the fluctuations in the field intensity, and consequentially increases the phase fluctuations, the phase squeezing increases the field intensity fluctuations, in contrast leading to a better defined phase. 
For instance, classical coherent states saturate the product of these two uncertainties, whereas the non-classicality in squeezed states is manifested if the variance of one field quadrature is below the level of the vacuum fluctuations \cite{walls1983squeezed}.
Squeezing of the field in HHG has been reported recently due to dipole moment correlations \cite{stammer2023entanglement} or when driven by squeezed light \cite{tzur2023generation}. 
An illustration of the aforementioned states in phase-space by means of their Wigner function can be seen in Fig.\ref{fig:HHG_wigner}.

Moreover, quantum optics allows us to consider driving fields beyond the classical perspective, and study the process of HHG driven by fields differing from the conventional assumption of coherent laser light was initiated in \cite{gorlach2023high}. In there, it was shown that light fields with a higher intensity fluctuation than classical coherent laser light, such as bright squeezed vacuum or thermal states, lead to an extended cut-off in the HHG spectrum. In contrast, intense photon number states have a well defined intensity and therefore have the same cut-off as the conventional classical coherent driving field. 
The well defined intensity of the photon number states has the consequence of an arbitrary phase and leads to vanishing mean electric field values at all times $E_{cl}(t) = 0$. 
Consequentially, this poses the question about the validity of the semi-classical perspective in such cases \cite{stammer2023limitations}, and the role of the optical phase of the driving field in the process of HHG was studied in \cite{stammer2023role}.
This ultimately raises the question of whether quantum optical coherence of the driving field, in terms of non-vanishing off-diagonal density matrix elements, is necessary to generate high-order harmonics. Indeed, it was recently shown in \cite{stammer2023role} that quantum optical coherence is not needed to drive the HHG process and that the generated harmonics are diagonal in their respective photon number basis. 
This has the intriguing consequence that inferring the coherence properties of the harmonics from the HHG spectrum alone is not justified. 

\subsubsection{Entanglement of the field modes}

Besides the Wigner function representation of a single field mode, there is a crucial aspect of quantum theory when considering more than a single mode. 
Considering all the field modes participating in the process of HHG we can study entanglement within the total quantum state (for a brief introduction to entanglement see section \ref{sec:2}). 
As shown in Eq.~\eqref{eq:product_state} the final state of all field modes is a product state, which holds when neglecting dipole moment correlations (see Eq.~\eqref{eq:Kraus_approx}). 
It was shown that this is related to the assumption of negligible ground state depletion of the electron in the case of moderate driving laser intensities \cite{stammer2022theory, stammer2023entanglement}.
However, as indicated in \cite{stammer2023quantum2} this approximation has the consequence of neglecting high order terms, such that the field operators of different modes do not mix. It was thus shown in \cite{stammer2023entanglement} that going beyond the assumption of negligible continuum state population leads to mode mixing, and thus, all field modes in the HHG process are, in general, entangled. 
Using the process of HHG, therefore, allows the generation of massive and bright entangled states using conditioning schemes \cite{stammer2022high} or resonant media \cite{yi2024generation}.

\begin{table*}[t]  
\caption{Overview on the recent findings for the full quantum optical description of the HHG process and its classification into classical (\textbf{x}) and quantum processes (\checkmark), or observations which are yet not classified. In case a classical analog exists, the respective counterpart is given.  }
\label{table:HHG}
  \centering
\begin{tabular}{ |p{4.6cm}||p{5.5cm}|p{5.8cm}| c | }
 \hline
 \multicolumn{4}{|c|}{\textbf{Quantum phenomena in HHG} } \\
 \hline
& Observable & Classical counterpart & Quantum \\
 \hline
 \hline
HHG emission \cite{lewenstein2021generation, rivera2022strong} & HHG spectrum & Dipole moment expectation value   & \textbf{x}  \\
\hline 
Field mode entanglement \cite{stammer2022high, stammer2022theory, stammer2023entanglement} & Correlation measurement (unknown) & Unknown & \textbf{?}\footnote{Although entanglement between the field modes is a feature of the full quantum optical description, the associated correlations in the observations can still have classical counterparts.} \\
\hline 
Mode squeezing \cite{stammer2023entanglement, tzur2023generation} & Homodyne detection & \textbf{x}  & \checkmark\footnote{Note that in \cite{tzur2023generation} only a fraction of the harmonic modes exhibit non-classical squeezing below the vacuum fluctuatons, and the residual harmonics are classical.} \\
\hline 
\multirow{2}{*}{Quantum optical coherence~\cite{stammer2023role} } & Spectrum \& Photon statistics  & $\rho_q = \frac{1}{2\pi } e^{- \frac{\abs{\chi_q}^2}{2}} \int_0^{2\pi} d\phi \dyad{\abs{\chi_q} e^{i \phi}} $  & \textbf{x}\footnote{The phase mixed coherent state $\rho_q$ gives rise to the same HHG spectrum and photon statistics as the pure coherent state $\ket{\chi_q}$ of equal amplitude. The mixed state $\rho_q$ can equivalently be expressed in diagonal form in the photon number basis $\rho_q = e^{- \abs{\chi_q}^2} \sum_n \frac{\abs{\chi_q}^{2n}}{n!} \dyad{n}$, and therefore does not exhibit quantum optical coherence~\cite{stammer2023role}.} \\ \cline{2-4}
& Interferometric measurements &  Unknown & \textbf{?}\footnote{Although the HHG spectrum and the photon statistics do not allow to infer on the quantum coherence properties of the harmonic radiation, it does not exclude possible interferometric measurements to reveal quantum coherence, i.e. the necessity of off-diagonal density matrix elements in the photon number basis to describe the phenomena.} \\
\hline 
\multirow{2}{*}{Non-Gaussian Wigner function } & Optical cat states $W(\beta) \le 0$ \cite{lewenstein2021generation} & \textbf{x} & \checkmark \\ \cline{2-4}
& Wigner function with $W(\beta) \ge 0$ \cite{pizzi2023light} & Mixed coherent state $\rho = \sum_i p_i \dyad{\alpha_i}$ & \textbf{x}\footnote{The fact that non-Gaussian Wigner functions refer to non-classical states only holds for pure quantum states \cite{hudson1974wigner}, leading to negativities in the Wigner function. } \\
\hline 
HHG for $E_{cl} (t) = 0$ \cite{stammer2023limitations, stammer2023role} & HHG spectrum & Driven by phase mixed coherent state  & \textbf{x}\footnote{The phase mixed coherent stat $\rho \propto \int_0^{2 \pi} d\phi \dyad{\abs{\alpha} e^{i \phi}} $ allows to generate high harmonics, which makes the assumption of non-vanishing electric fields unnecessary to describe the HHG process \cite{stammer2023role}. Further, it was recently shown that HHG in the semi-classical description also occurs for vanishing classical electric fields \cite{weber2023quantum}. However, the subtle difference here is that for the driving fields considered in \cite{stammer2023limitations, stammer2023role}, the electric field vanishes at all times.} \\
\hline 
Extended HHG cut-off  \cite{gorlach2023high} & HHG spectrum & Driven by classical fluctuating fields  & \textbf{x}\footnote{The extension of the HHG cut-off was shown to occur also for classical fields exhibiting large fluctuations in the field intensity, such as thermal states~\cite{gorlach2023high}.} \\
\hline 
Analog Simulation of HHG \cite{arguello2024analog} & Absorption image for HHG spectrum & HHG spectrum & \textbf{x}\footnote{Although quantum simulators \cite{buluta2009quantum} are a versatile platform to study quantum phenomena, the approached used in \cite{arguello2024analog} to simulate the HHG spectrum is based on classical interactions. } \\
\hline

\end{tabular}

\end{table*}

\subsection{Above threshold ionization}

For the process of above-threshold ionization (ATI), the semi-classical description has thus far focused on the photoionization process alone. However, due to the inherent limitations of such a classical description of the light field, it is easily overlooked that the photoionization process can be accompanied by the emission of radiation. In the following, we highlight the insights obtained when analyzing the ATI process in light of a full quantum optical perspective. 

\subsubsection{Quantum field coupled to ATI electrons}

The photoionization mechanism in ATI can be well described by considering classical driving fields, while the radiation properties are often neglected in the process. 
However, recent advances in the full quantum optical description of ATI \cite{stammer2023quantum} have revealed that the photoionization process is accompanied by noteworthy consequences for the light field. 
For instance, it was shown that the ATI process is accompanied by the emission of radiation and that the driving laser field can experience variable changes depending on the emission direction of the photoelectron \cite{stammer2023quantum}.
Further, it was shown that quantizing the radiation field allows to study the role of entanglement between the emitted electron and the simultaneously emitted light field \cite{rivera2022light}. This allows to study hybrid light-matter entangled systems, which can potentially be used for quantum state engineering of light.  
In \cite{milovsevic2023quantum} it was shown, that the light emission during ATI can be controlled by varying the laser intensity, wavelength and by considering different driving field polarizations.

\subsubsection{ATI driven by non-classical light}

As was shown for the process of HHG (see section \ref{sec:HHG_properties}), driving fields beyond the classical perspective leads to new observations. 
This is also the case for ATI as shown in \cite{even2023photon}, in which a bright squeezed state is considered to induce the photoionization mechanism, with profound consequences on the dynamical properties of the driven electron.
This has measurable consequences on the photoelectron momentum distribution of the ionized electron \cite{fang2023strong}.
Furthermore, the observed extension of the HHG-cutoff for driving fields with increased intensity fluctuations, the same occurs in photoionization. It was shown in \cite{wang2023high} that the photoelectron energy distribution experiences a similar extension of the cutoff energy.

\section{\label{sec:2}Entanglement phenomena in attoscience}

Entanglement \cite{Horodecki2009} describes an inherent quantum phenomena where the components within a quantum system become interconnected in such a way that the individual subsystem has no definite description. 
Consequently, the state of a multipartite quantum system cannot be adequately represented by a single product of wave functions $\ket{\psi_i}$ describing the individual parts, i.e. $\ket{\Psi} \neq \bigotimes_i \ket{\psi_i}$. 
In particular, the corresponding quantum correlations can reveal stronger correlations within the composite system than classically allowed, and contradict the classical assumption that any physical object possesses predetermined and individual properties. Quantum entanglement is a fundamental component of quantum theory and has been the subject of extensive research and fascination in the realm of physics.
In the following, we delve into the significance of quantum entanglement in photoionization or photoionization-induced processes. These processes form the foundation for a large variety of spectroscopic and imaging techniques within the field of attosecond science.

\begin{table*}[t]
\caption{The table summarizes the different entanglement measures together with the advantages and disadvantages as well as how they have been used in the context of attosecond science. }
\label{table:entanglement}
  \centering
\begin{tabular}{|p{4.5cm}|p{5.0cm}|p{4.0cm}|p{4.5cm}| }
\hline
\multicolumn{4}{|c|}{\textbf{Entanglement measure and witness in attoscience} } \\
\hline
\textbf{Entanglement Measure}& \textbf{Relation to Entanglement} & \textbf{Applications in Attoscience}  & \textbf{Pros and Cons}\\
\hline
\hline
\textbf{Purity:}   $$P = \Tr (\rho_A^2)\leq 1,$$ where $\rho_A = \Tr_B (\dyad{\psi_{AB}})$ is the reduced density matrix & $1/d \le P \le 1$, with $d$ the dimension of the Hilbert space. For separable states $P=1$ and maximally entangled states $P = 1/d$. & Measure of the ion-photoelectron entanglement in photoinization process \cite{vrakking2021control,Busto2022,Vrakking2022,Laurell2022}
&\begin{itemize}[noitemsep,nolistsep]
    \item entanglement measure for pure states
    \item non-trivial extension to entangled mixed states
\end{itemize}
\\
\hline
\textbf{Entanglement entropy:} $$S(\rho_A) = - \Tr (\rho_A \log \rho_A ),$$ where $\rho_A$ is the reduced density matrix & $S(\rho_A) = 0$ for pure states $\rho_A$ and therefore separable $\dyad{\psi_{AB}}$ and $S(\rho_A)$ maximal for maximally mixed $\rho_A$ and thus maximally entangled $\dyad{\psi_{AB}}$ 
& Measure of the ion-photoelectron entanglement in photoinization process  \cite{Ruberti2019,Nabekawa2023}
& 
\begin{itemize}[noitemsep,nolistsep]
    \item clear interpretation in terms of entropy and information theory
    \item non-trivial extension to entangled mixed states
\end{itemize}
\\
\hline
\textbf{Logarithmic negativity:} $$\mathrm{LN}(\rho) =\log \| \rho^\Gamma \|_1$$ where $\rho^\Gamma$ is the partial transpose of $\rho$ and $\| A \|_1 = \Tr[\sqrt{A^\dagger A}]$ denotes the trace norm & Uses Positive Partial Transpose criterion: applies the partial transpose to the system and check if $\rho^\Gamma = (\mathds{1}_A \otimes T_B) [\rho] \ge 0$ positive semi-definite (equivalent to positive eigenvalues). But not a sufficient condition for entanglement in arbitrary dimensions & Quantifies entanglement between the OAM of the two photoelectrons in NSDI \cite{maxwell2022entanglement} & \begin{itemize}[noitemsep,nolistsep]
    \item simple computation
    \item directly applicable to mixed states
    \item can not detect entanglement in arbitrary dimensions of the subsystems, i.e. state can be entanglement but still $LN(\rho) =0$
\end{itemize} \\
\hline
\textbf{Entanglement witness} $\mathcal{W}$ is an operator that is positive on all
separable states
& Associated with observables whose expectation values are negative for entangled states, i.e. $$\Tr (\mathcal{W}\rho)<0 $$
Distinguish a subset of entangled states as non-separable.& Quantifies entanglement between the OAM of the two photoelectrons in NSDI \cite{maxwell2022entanglement} & \begin{itemize}[noitemsep,nolistsep]
    \item directly applicable to mixed states
    \item versatile, can be tailored to specific system under consideration
\end{itemize}
    
\\
\hline
\end{tabular}

\end{table*}
\subsection{Ion-photoelectron entanglement}
Photoionization involves the disintegration of a complex many-electron quantum system, like an atom or molecule, into an emitted photoelectron drifting away and the remaining parent ion. The induced dynamics of the resulting ionic or photoelectronic subsystem is often the focus in attosecond experiments. These experiments often rely on interference effects, and the measurements typically focus on probing the ultrafast dynamics either within the ion \cite{SansoneNature2010,GoulielmakisNature10,CalegariScience14} or photoelectron system \cite{PaulScience01,Kienberger2002}.\\
In most cases, theoretical investigations of photoionization-induced processes treat the parent ion and photoelectron separately, overlooking their possible quantum entanglement. For instance, in studies concerning ultrafast hole dynamics and the associated charge migration in polyatomic molecules \cite{CederbaumCPL99,Despre2015,Du2019,MatselyukhNatPhys2022,FolorunsoJPCA2023}, a sudden coherent wave packet formation of the parent ion is assumed yielding a well-defined initial state.
However, the question how such a coherent state is initially generated remains. Similarly, in studies centered on the emitted photoelectron, the quantum correlations of the combined system have largely been disregarded \cite{Guillemin2015}.\\
Nevertheless, quantum entanglement arising between the parent ion and photoelectron during attosecond photoionization can significantly influence the entire quantum system. Particularly, entanglement constrains coherence properties, potentially hindering the observation of delay-dependent observables sensitive to these coherences \cite{Carlstrom2018,RubertiPCCP2022}. Although the precise role of ion-photoelectron entanglement in attosecond science remains largely unexplored, recent experimental and theoretical studies have increasingly focused on understanding the influence of quantum entanglement on experimental observations.\\
For example, pioneering experimental research on the evolution of an electron hole in ionized $\mathrm{Kr}$ atoms \cite{GoulielmakisNature10} already indicated how the entanglement with the accompanying photoelectron might restrict electronic coherence \cite{Smirnova2010}, further supported by theoretical work on similar systems \cite{Pabst2011,Karamatskou2020}. Furthermore, entanglement acting as a limiting factor for vibrational and electronic coherence of ionized $\mathrm{H_2}$ molecules was demonstrated by several papers \cite{vrakking2021control,Vrakking2022,koll2022experimental,Nishi2019,Nabekawa2023}. The measure used to quantify the degree of entanglement has been the purity of the reduced density matrix of one subsystem (see Table II). In \cite{vrakking2021control}, the purity of the reduced density matrices of the ion and photoelectron was used to show how the degree of entanglement can be controlled by changing the time delay between two ionizing extreme ultraviolet (XUV) pulses. Fig.~\ref{fig:Fig3} (a) shows the purity as a function of the XUV-XUV time delay for different pulse duration. This was followed by the experiment \cite{koll2022experimental} showing how the vibrational coherence of the $\mathrm{H_2}^+$ ions can be controlled by varying this time delay. Additionally, it has been explored how using a chirped XUV pulse can also affect the degree of entanglement \cite{Vrakking2022}. Fig~\ref{fig:Fig3}(b) presents the purity of the ion-photoelectron density matrix as a function of the chirped parameter of the XUV pulse.
Overall, these studies \cite{vrakking2021control,koll2022experimental,Vrakking2022} extensively discuss the inherent relation between quantum entanglement and coherence, and propose the potential manipulation of these quantities by adjusting the spectral attributes of the photoionizing pulse.
The characteristics of the ionizing pulse, e.g. pulse duration or mean photon energy will dictate the kinetic energy of the photoelectron and thus its proximity to the parent ion that will result in different interchannel couplings to the ionic states and therefore the entanglement \cite{Pabst2011,RubertiPCCP2022}.

Entanglement naturally arises here since different combinations of rovibrational states of the parent ion and angular momentum states of the ejected electron are possible, such that the total wavefunction represents an entangled state.
To that end, a witness and quantification of this entanglement is of current interest. Since the underlying process is dynamical, its measurement will at first require the identification of appropriate observables that are easy to detect and which allow to infer about entanglement. 
A possible proof of quantum entanglement would be a (loophole free) test of the violation of Bell inequalities \cite{Bell1964}, which in the case of measuring ion-photoelectron entanglement, means detecting correlations in the measurement of two non-commuting observables for the ion and the photoelectron. Such a test has been recently proposed in the context of attosecond science \cite{ruberti2023bell}.
The authors theoretically reported a violation of Bell inequalities in the photoionization of argon atoms, with the degree of freedom encoded in the spin angular momentum of the photoelectron. The simulations, based on first principles, reported on correlations in the measurements of the observables which violate the Bell inequality; while the challenge lies in identifying the appropriate pairs of non-commuting observables. On the experimental side, a considerable promise to quantify entanglement due to spin-orbit interactions was demonstrated \cite{Laurell2022}. Within this work a continuous-variable quantum state tomography of the photoelectron was used -- a protocol that is commonly used in quantum optics to tomographically reconstruct the quantum state of a system, which they further validated with density matrix calculations.

\subsection{Nuclear-electronic entanglement}
In the field of molecular dynamics, nuclear and electronic motion are generally associated with different timescales justifying the conventional practice of separating the slowly moving nuclei from the fast moving electrons building the foundation of the Born-Oppenheimer (BO) approximation. This approximation serves as fundamental framework, particularly in visualizing potential pathways of photochemical and photophysical processes. Although simulations involving excited electronic states often require to go beyond the traditional BO picture by incorporating non-adiabatic coupling terms (if electronic states cross each other), the initial step of partitioning the molecular system into distinct, independent electronic and nuclear degrees of freedom persists.

In investigations primarily centered on nuclear dynamics, this assumption manifests in the concept of a potential energy surface (PES). These surfaces represent solutions to the electronic problem alone and depend solely on the nuclear position in a parametric manner. Subsequently, the nuclear problem is solved, yielding the well-established depiction of nuclear wavepackets moving on PESs (within and beyond the BO approximation). However, this sequential resolution of the electronic and nuclear problems relies on the existence of two disentangled sets of degrees of freedom.

In many theoretical inquiries into purely electronic phenomena in molecules, such as hole migration, electron dynamics is often simulated at fixed nuclear geometries \cite{CederbaumCPL99,RemaclePNAS06,Yong2022}. This practice again overlooks the potential entanglement between the nuclear and electronic degrees of freedom. Nevertheless, several studies have already indicated a high correlation between electronic and nuclear dynamics, suggesting the likelihood of electronic decoherence or even recoherence due to nuclear motion \cite{VacherPRL17,Arnold2017,ArnoldPRL18,JiaJPCL19,DeyPRL22}. Specifically, the dependence of quantum coherence on electronic-nuclear entanglement created by non-adiabatic coupling has been demonstarted in \cite{Vatasescu2013,Vatasescu2015}. While non-adiabatic couplings generally lead to both a breakdown of the BO approximation and to entanglement, the actual degree of electron-nuclear entanglement and the degree of validity of the BO approximation are unrelated \cite{Izmaylov2017}. Moreover, recent work \cite{Blavier2022} delves into the time evolution of entanglement, utilizing a matrix representation of the wave functions for the entangled systems. A singular value decomposition of this matrix at each time will result in singular values, whose time evolution will govern the time evolution of entanglement. Their findings not only confirm the influence of non-adiabatic couplings on molecular entanglement but also highlight the interaction with the exciting pulse as a major contributing factor.

Thus, even under the assumption of complete unentangled parent ion and photoelectron, with the ionic subsystem exhibiting full coherence, internal entanglement may persist between the molecular electronic and nuclear degrees of freedom \cite{RubertiPCCP2022}. This internal entanglement has implications for electronic coherence and consequently impacts the outcomes of experiments in the field of attoscience.

\subsection{Electron-electron entanglement}

\begin{figure}[t]
    \centering
	\includegraphics[width=1\columnwidth]{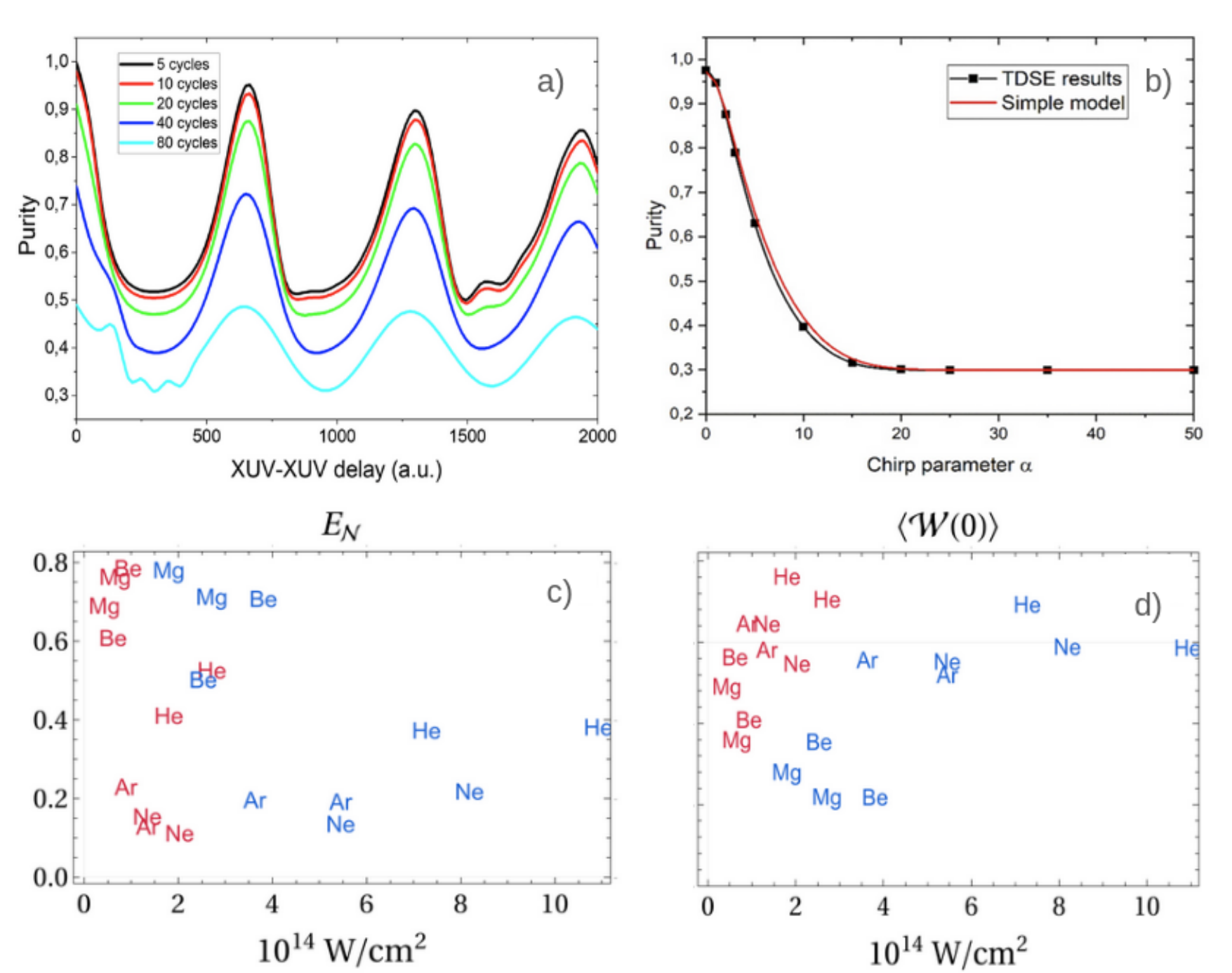}
	\caption{The figure shows different entanglement measures and their applications in attoscience.  Panel (a) shows the purity of the reduced density matrix of the ion-photoelectron system as a function of the time delay between two XUV pulses for different pulse durations. 
    The theoretical predictions show how, by tuning the time delay, we can control the degree of entanglement from low (high purity) to high (small purity), and thus have a high and low degree of vibrational coherence, respectively.
    Panel (b), adapted from \cite{Vrakking2022}, shows the dependence of the purity of the ion-photoelectron density matrix as a function of the chirp of the XUV pulse. The figure shows theoretical results obtained from the solution of the time-dependent Schr\"odinger equation indicating the transition from a pure to a mixed state and can be understood as the consequence of the ion-photoelectron entanglement. Panels (c) and (d), adapted from \cite{maxwell2022entanglement}, present the logarithmic negativity $E_\mathcal{N}$ and the average value of the entanglement witness $\langle\mathcal{W}\rangle$, respectively as constructed in \cite{maxwell2022entanglement}. The figure shows an extensive search over different targets and field intensities, the blue text corresponds to $\lambda=400 \; \rm nm$, and the red text to $\lambda=800 \; \rm nm$. Panel (a) reprinted with permission from \cite{vrakking2021control}. Copyright 2021 by the American Physical Society. Panel (b) reprinted from \cite{Vrakking2022} under the premises of the Creative Commons Attribution 4.0 International License. Panels (c) and (d) reprinted from \cite{maxwell2022entanglement} under the premises of the Creative Commons Attribution 4.0 International License \url{https://creativecommons.org/licenses/by/4.0/}
} 
      \label{fig:Fig3}
\end{figure}

Electron-electron correlation has been observed from the early days of attosecond science \cite{l1983multiply} in the context of laser-induced double and multiple ionization. In particular, the process of non-sequential double ionization (NSDI) has been extensively studied, revealing the role of electron-electron correlation in the laser-induced dynamics \cite{WeberNat00,becker2012theories} where the inter-electronic Coulomb interaction governs the entanglement \cite{Akoury2007}. 

The theoretical modeling of this phenomena has been controversial. Although classical models have been quite successful in modeling NSDI \cite{HoPRL05}, it has been debated whether we are in the presence of classical or quantum electron-electron correlation. Several studies have revealed the onset of quantum interference between the two photoelectrons \cite{hao2014quantum,maxwell2016controlling,quan2017quantum}. However, the study of electron-electron entanglement has not been the center of attention. One important reason is that typically measuring entanglement will require evaluating the continuous variable density matrix or entanglement measures like the purity or von Neumann entropy, which are very challenging to compute (see Table II for a summary of the different measures of entanglement, the advantages and disadvantages, and the applications in attoscience). A breakthrough in this direction has been made in \cite{maxwell2022entanglement}. They demonstrated the use of orbital angular momentum (OAM) to quantify and measure entanglement in NSDI. Here, entanglement arises due to the shared angular momentum during recollision with subsequent ionization mechanism. The use of the OAM significantly reduced the computational cost of evaluating the density matrix and allowed a simple evaluation of the logarithmic negativity and entanglement witnesses, exploring a wide range of targets and field intensities as shown in Fig.~\ref{fig:Fig3} panels (c) and (d). In particular, the entanglement witness can be decomposed into local measurements, which has a significant impact on experimental entanglement detection. 
Thus, in NSDI, which provides a direct manifestation of electron-electron correlation, the OAM entanglement of photoelectrons can reveal its fundamental non-classical nature. A question that remains open is how we can measure the OAM of the electrons. In the past decade, new techniques have become available to measure the OAM for electron vortex beams \cite{harvey2017stern,noguchi2019efficient}. As suggested in \cite{maxwell2022entanglement}, adding some of these techniques to the typical reaction microscope, which is already suitable for studying the many-particle entangled dynamics \cite{schmidt2021coltrims}, could be one of the directions to explore. Some challenges will arise using alkali metals, and further theoretical studies should be performed to enhance entanglement in noble gas targets.

\section{\label{sec:conclusion}Outlook}

In this Perspective, we have presented an overview of quantum phenomena in attosecond physics and particularly highlighted the emerging full quantum descriptions and the increasing interest in entanglement in attosecond processes.
We can see that this rapidly growing field of quantum phenomena in attosecond science provides new insights into the underlying dynamics and its consequences to experimental observations. 
Moreover, this emerging field still provides promising new perspectives for processes on the attosecond timescale and can open many different research directions in this field. 
In the following we present a list of potential future research directions to further emphasize on the deep connection between quantum phenomena and attosecond physics with the use for quantum information science \cite{lewenstein2022attosecond}.

\begin{itemize}
    \item  \textbf{Non-classical field properties:} As illustrated in Sec. \ref{sec:1} the full quantum optical description of attosecond processes has enabled to consider field observables beyond the classical realm. Investigating novel field properties, inaccessible before, is far from being uncovered and many different perspectives can be considered. In particular, genuine quantum features such as entanglement and squeezing of the optical field \cite{stammer2023entanglement, tzur2023generation} or driving the process by non-classical light \cite{gorlach2023high, stammer2023limitations, stammer2023role, tzur2023generation} are only being considered very recently. Properties such as the photon statistics or quantum noise as well as field correlation functions are only a fraction of what can be explored within the Hilbert space of the optical field modes inaccessible before.  

    \item \textbf{Field-field entanglement:} Interestingly, the derivation of the entangled field state in \cite{stammer2023entanglement} does not rely on specific strong field dynamics and is therefore generic to all parametric processes. We thus expect, that this opens a new avenue for studying and engineering massive entangled states of light when considering that Ref. \cite{pizzi2023light} indicated that material correlations can be imprinted in the field properties. This might allow to probe topological materials \cite{baldelli2022detecting, bera2023topological} or quantum phase transitions \cite{alcala2022high} using the process of HHG. 

    \item \textbf{Quantum state engineering of light:} The quantum optical description of HHG opened the avenue to use strong laser driven processes as a novel platform for quantum state engineering of light \cite{stammer2023quantum, bhattacharya2023strong}. 
    Further understanding the role of quantum optical coherence \cite{stammer2023role}, the importance of the material correlations \cite{pizzi2023light, stammer2023entanglement} as well as bringing these approaches to different targets such as molecular \cite{rivera2023quantum_molecule}, plasma \cite{lamprou2021plasma} or solid state systems \cite{rivera2024nonclassical, rivera2022quantum_wannierbloch, gonoskov2022nonclassical} is of current interest. 
    Further controlling the properties of the generated non-classical field states is of great importance and extending its range of application to non-linear optics \cite{lamprou2023nonlinear} or quantum metrology \cite{martos2023metrological} is of current interest. 

    \item \textbf{Influence of entanglement on electronic coherence:} In atomic or molecular photoionization, ion-photoelectron entanglement limits the electronic coherence in the ionic system \cite{GoulielmakisNature10,koll2022experimental}. The electronic coherence further depends on the spectral characteristics of the ionizing pulse \cite{Pabst2011} and any internal entanglement between the electronic and nuclear degrees of freedom in the ionic system \cite{Blavier2022}. Understanding the role of entanglement is of particular interest in charge migration studies to achieve charge-directed reactivity. Search for techniques to quantify the entanglement is currently in progress, with a considerable promise shown by the recent application of quantum state tomography in attoscience \cite{Laurell2022}. 

    \item \textbf{Entanglement in attosecond science:} An alternative route to treat the entanglement problem is the Bohmian framework \cite{pladevall2019applied}. This formulation offers a natural playground to distinguish between classical and quantum phenomena, and entanglement has already been addressed in several contexts. The applications of this theory to model entanglement in strong-field-induced phenomena are yet to be explored. The initial steps have been taken in \cite{christov2019phase} in the context of NSDI. Recently, an implementation of Bohmian mechanics using the conditional wave function method has been applied to model entanglement in multi-particle bosonic systems \cite{elsayed2018entangled}. The extension of such a model to multi-particle fermionic systems represents a promising alternative to understanding electron-electron entanglement in double and multi-ionization processes.

    \item \textbf{Conceive novel experiments:} For all of the aforementioned cases, it is inevitable to conceive novel experiments. On the one hand, to measure the previously unexpected field properties as well as to witness the proposed entanglement in attosecond processes \cite{koll2022experimental}. 
    This also includes techniques for the reconstruction of quantum states of the field or the electronic wavefunctions, which are currently limited to specific cases. 

    \item \textbf{Bringing attosecond science towards optical quantum technologies:} With the flourishing connection between quantum information science and attosecond physics, it seems likely that this symbiosis leads to the implementation of optical quantum technologies \cite{lewenstein2022attosecond}. 
    This can emerge from the recently developed quantum state engineering of light approaches or by using entanglement or quantum coherence as a resource for technologies.

\end{itemize}

\begin{acknowledgments}

The authors thank the organizers of the `Quantum Battles in Attoscience 2023' conference for the opportunity and support during the battle preparation.
L.C.R. acknowledges financial support from the AQuA DIP project grant number EP/J019143/1.
A.F. acknowledges financial support from the Cluster of Excellence 'CUI: Advanced Imaging of Matter' of the Deutsche Forschungsgemeinschaft (DFG) - EXC 2056 - project ID 390715994, from the International Max Planck Graduate School for Ultrafast imaging $\&$ Structural Dynamics (IMPRS-UFAST) and from the Christiane-N\"usslein-Vollhard-Foundation.
D.D. acknowledges funding from the National Science Foundation under Grant No. CHE-2347622 and the European Union's Horizon 2020 research and innovation programme under the Marie Sklodowska-Curie Grant agreement No. 892554.
P.S. acknowledges funding from the European Union’s Horizon 2020 research and innovation programme under the Marie Skłodowska-Curie grant agreement No 847517. 
ICFO group acknowledges support from:
ERC AdG NOQIA; MCIN/AEI (PGC2018-0910.13039/501100011033, CEX2019-000910-S/10.13039/501100011033, Plan National FIDEUA PID2019-106901GB-I00, Plan National STAMEENA PID2022-139099NB-I00 project funded by MCIN/AEI/10.13039/501100011033 and by the “European Union NextGenerationEU/PRTR" (PRTR-C17.I1), FPI); QUANTERA MAQS PCI2019-111828-2); QUANTERA DYNAMITE PCI2022-132919 (QuantERA II Programme co-funded by European Union’s Horizon 2020 program under Grant Agreement No 101017733), Ministry for Digital Transformation and of Civil Service of the Spanish Government through the QUANTUM ENIA project call - Quantum Spain project, and by the European Union through the Recovery, Transformation and Resilience Plan - NextGenerationEU within the framework of the Digital Spain 2026 Agenda; Fundació Cellex; Fundació Mir-Puig; Generalitat de Catalunya (European Social Fund FEDER and CERCA program, AGAUR Grant No. 2021 SGR 01452, QuantumCAT \ U16-011424, co-funded by ERDF Operational Program of Catalonia 2014-2020); Barcelona Supercomputing Center MareNostrum (FI-2023-1-0013); EU Quantum Flagship (PASQuanS2.1, 101113690, funded by the European Union. Views and opinions expressed are however those of the author(s) only and do not necessarily reflect those of the European Union or the European Commission. Neither the European Union nor the granting authority can be held responsible for them); EU Horizon 2020 FET-OPEN OPTOlogic (Grant No 899794); EU Horizon Europe Program (Grant Agreement 101080086 — NeQST), results incorporated in this standard have received funding from the European Innovation Council and SMEs Executive Agency under the European Union’s Horizon Europe programme)., ICFO Internal “QuantumGaudi” project; European Union’s Horizon 2020 program under the Marie Sklodowska-Curie grant agreement No 847648; “La Caixa” Junior Leaders fellowships, La Caixa” Foundation (ID 100010434): CF/BQ/PR23/11980043. Views and opinions expressed are, however, those of the author(s) only and do not necessarily reflect those of the European Union, European Commission, European Climate, Infrastructure and Environment Executive Agency (CINEA), or any other granting authority. Neither the European Union nor any granting authority can be held responsible for them.

\end{acknowledgments}

\section*{Author contributions}
The authors contributed equally to the writing of the manuscript.

\section*{Competing interests}
The authors declare no competing interests.

\bibliographystyle{unsrt}
\bibliography{references}{}

\begin{thebibliography}{100}
\bibitem{footnote1}
Note that this is not the only non-classical witness using phase-space
  representations~\cite{mandel1986non}, and also states with $W(\beta) \ge 0$
  can be ne non-classical such as squeezed states~\cite{loudon1987squeezed}.
\end{thebibliography}

\end{document}